\documentclass[aps,twocolumn,prl,showpacs,showkeys,preprintnumbers,superscriptaddress,nobibnotes,floatfix,nofootinbib]{revtex4-2}

\usepackage{amsmath}
\usepackage{amsfonts}
\usepackage{amssymb}
\usepackage{mathrsfs}
\usepackage{bm} 
\usepackage{bbm} 
\usepackage{slashed} 
\usepackage{physics} 
\usepackage{esvect} 
\usepackage[version=4]{mhchem}
\usepackage{xcolor} 
\usepackage{graphicx} %
\usepackage{array} 
\usepackage{url} %
\usepackage{amsthm} 
\usepackage{enumerate} 
\usepackage{enumitem} %
\usepackage{booktabs} 
\usepackage{siunitx} 
\usepackage{mdframed} %
\usepackage{tcolorbox} %
\usepackage{fontawesome5} 
\usepackage{braket} 
\usepackage{setspace} %
\usepackage{multirow} %
\usepackage{upgreek} 
\usepackage{xspace} %
\usepackage[normalem]{ulem} 
\usepackage{array}
\usepackage{natbib} 
\usepackage{hyperref} %
\hypersetup{
    colorlinks=true,
    linkcolor=red,
    citecolor=blue,
    filecolor=magenta,
    urlcolor=blue,
    breaklinks=true
}
\usepackage[all]{hypcap} 
\usepackage[capitalise]{cleveref} %

\setlist[description]{leftmargin=\parindent,labelindent=\parindent} %

\newcommand{\nl}{\nonumber\\}

\newcommand{\MyTr}[1]{\operatorname{Tr}\left[#1\right]}

\newcommand{\SPSD}[2][]{{\mathcal{S}^{#1}_{#2}}}
\newcommand{\BPSD}[2][]{{\mathcal{B}^{#1}_{#2}}}

\newcommand{\average}[1]{\left\langle #1 \right\rangle}

\DeclareMathOperator\Var{Var}

\newcommand{\prlsection}[2]{{\it\textbf{#1}{#2}}---}

\begin{document}

\preprint{KEK-QUP-2025-0023, KEK-TH-2769}

\title{Beyond Qubits: Multilevel Quantum Sensing for Dark Matter}

\author{Xiaolin Ma}  
\altaffiliation{\href{mailto:xlmphy@post.kek.jp}{xlmphy@post.kek.jp, corresponding}} 
\affiliation{International Center for Quantum-field Measurement Systems for Studies of the Universe and Particles (QUP, WPI),
High Energy Accelerator Research Organization (KEK), Oho 1-1, Tsukuba, Ibaraki 305-0801, Japan}

\author{Volodymyr Takhistov}  
\altaffiliation{\href{mailto:vtakhist@post.kek.jp}{vtakhist@post.kek.jp, corresponding}}
\affiliation{International Center for Quantum-field Measurement Systems for Studies of the Universe and Particles (QUP, WPI),
High Energy Accelerator Research Organization (KEK), Oho 1-1, Tsukuba, Ibaraki 305-0801, Japan}
\affiliation{Theory Center, Institute of Particle and Nuclear Studies (IPNS), High Energy Accelerator Research Organization (KEK), \\ Tsukuba 305-0801, Japan
}
\affiliation{Graduate University for Advanced Studies (SOKENDAI), 
1-1 Oho, Tsukuba, Ibaraki 305-0801, Japan}
\affiliation{Kavli Institute for the Physics and Mathematics of the Universe (WPI), UTIAS, \\The University of Tokyo, Kashiwa, Chiba 277-8583, Japan}

\author{Norikazu Mizuochi}   
\affiliation{Institute for Chemical Research, Kyoto University, Gokasho, Uji-city, Kyoto 611-0011, Japan}
\affiliation{Center for Spintronics Research Network, Kyoto University, Uji, Kyoto 611-0011, Japan}
\affiliation{International Center for Quantum-field Measurement Systems for Studies of the Universe and Particles (QUP, WPI),
High Energy Accelerator Research Organization (KEK), Oho 1-1, Tsukuba, Ibaraki 305-0801, Japan}

\author{Ernst David Herbschleb}   
\affiliation{Institute for Chemical Research, Kyoto University, Gokasho, Uji-city, Kyoto 611-0011, Japan}

\begin{abstract}
Quantum sensing with qubits has advanced fundamental physics searches, but higher dimensional systems offer untapped potential. We present a universal qutrit framework that yields a sequence-independent fourfold increase in quantum Fisher information and a twofold gain in sensitivity. In ultralight dark matter searches, spin-1 NV-center qutrits can enhance the axion-electron coupling reach by an order of magnitude beyond qubits. This principle applies broadly to multilevel quantum systems including superconducting, neutral atom and trapped-ion qutrits, establishing higher dimensional sensing as a powerful tool for probing new physics.
\end{abstract}

\maketitle

\prlsection{Introduction}{.}
Quantum sensing has emerged as powerful approach harnessing quantum systems for probing extremely weak signals with unprecedented precision~\cite{Degen:2016pxo}. Applications span fields from condensed matter and chemistry to sensitive tests of fundamental interactions. Many sensing techniques rely on control of qubit two quantum level systems whose coherence establishes the sensitivity with a variety of realizations such as superconducting, neutral and trapped-ion qubits.

However, many quantum systems possess richer internal structures, such as spin-1 states $|0\rangle$ and $|\pm1\rangle$ that enable three level qutrit superpositions, capable of encoding additional information. More generally, qudits have attracted broad interest in quantum information science offering potential advantages such as in error correction \cite{Campbell:2012olh,Campbell:2014dok} and  quantum communication \cite{Bouchard:2016zre}. Controllable superconducting \cite{Yurtalan:2020tlg,Goss:2022bqd}, neutral atom~\cite{Lindon:2022ekh} and trapped-ion  \cite{Ringbauer:2021lhi} qutrits have recently been experimentally achieved. Among sensing techniques, nitrogen-vacancy (NV) centers in diamond stand out as an established quantum technology providing precision nT magnetic field sensitivity for single NV centers, nano-scale spatial resolution, and room-temperature operation \cite{Degen:2016pxo,Barry:2019sdg,Herbschleb:2019zlq}. Recently, controllable NV qutrits have been realized~\cite{Barry:2023hon,zhou2017holonomic}.
While qutrits and qudits have been studied in quantum information science, their fundamental physics potential remains poorly explored.

In this Letter, we show that exploiting the intrinsic higher dimensional structure of quantum systems enables a universal and sequence-independent enhancement in sensitivity for fundamental physics searches, particularly those targeting dark matter (DM). DM is the dominant form of matter in the Universe, whose nature remains one of the most profound open questions in science. Ultralight bosons with masses $m_a \ll \mathcal{O}(1)$~eV, such as axions and axion-like particles (ALP), are well motivated candidates from fundamental theory~\cite{Peccei:1977hh,  Weinberg:1977ma, Wilczek:1977pj, Vafa:1984xg,Essig:2013lka,ParticleDataGroup:2022pth,Dine:1982ah}. A variety of quantum sensing approaches based on qubits, ranging from superconducting~\cite{Dixit:2020ymh,Ikeda:2021mlv,Agrawal:2023umy,Chen:2023swh,Chen:2024aya,Braggio:2024xed}  to trapped-ion systems~\cite{Ito:2023zhp}, have been recently proposed for ultralight DM detection.
We explicitly demonstrate that NV center qutrits amplify DM-induced signals beyond recently proposed NV qubit setups~\cite{Chigusa:2023roq,Chigusa:2024psk}, showing realistic improvements in axion-electron coupling sensitivity. The principle we introduce provides a general framework applicable to a broad class of  sensing platforms.

\prlsection{Quantum Fisher information enhancement}{.}
Qutrits with three quantum states can outperform qubits enhancing the detection of weak fields.
The intrinsic sensitivity advantage of employing qutrits can be demonstrated through the quantum Fisher information (QFI) $F_q$ that measures sensing precision, setting the ultimate limit for estimating a parameter $\theta$ encoded in a quantum state via the Cramer-Rao bound $(\Delta\theta)^2 \ge 1/F_q$. A larger QFI corresponds to a greater achievable sensitivity~\cite{Liu:2019xfr,Cowan:2010js}.

For a quantum state $|\psi(0)\rangle$ in a system evolving under an external field that couples linearly to the spin operator $\hat S_z$, the accumulated phase $\theta$ leads to state
\begin{align}
    |\psi\rangle=e^{i \hat{S}_z \theta} |\psi(0)\rangle.
    \label{eq:perturbe_state}
\end{align}
The phase $\theta$ reflects the time-integrated effect of the field 
$\theta=\gamma_e \int \mathcal{B}(t) y(t) dt$, where $\gamma_e\simeq e/m_e$ is the electron gyromagnetic ratio for electron of mass $m_e$ and $\mathcal{B}(t)$ denotes the effective and possibly novel magnetic or pseudo-magnetic field and $y(t)=\pm1$ is the control or modulation function defined by the sensing pulse sequence.
For such a pure state undergoing a unitary evolution, the QFI is given by \cite{Liu:2019xfr}
\begin{equation}
    F_q(\theta) =4(\langle \partial_\theta \psi|\partial_\theta \psi \rangle -|\langle \partial_\theta \psi|\psi \rangle |^2) \nl
     =4\langle\Delta \hat{S}_z^2\rangle,
\end{equation}
where $\langle\Delta \hat{S}_z^2\rangle = \langle\hat{S}_z^2\rangle - \langle\hat{S}_z\rangle^2$ is the variance of $\hat{S}_z$ in the initial state $|\psi(0)\rangle$.

\begin{figure}[t]
    \centering
    \includegraphics[width=1\linewidth]{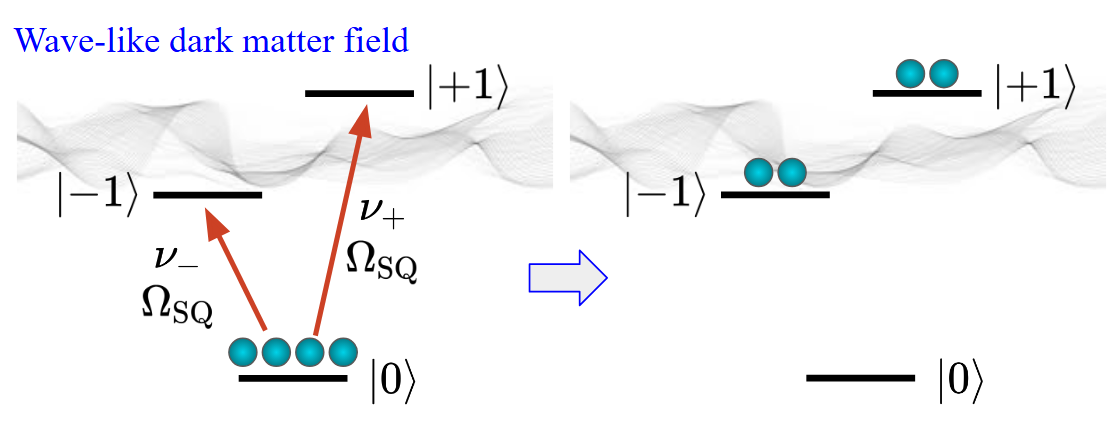} 
    \caption{Schematic illustration of qutrit DM sensing proposed in this work.
A wave-like DM field couples a spin triplet, driving simultaneous transitions between the $\ket{0}$ and $\ket{\pm 1}$ states with Rabi frequency $\Omega_{\mathrm{SQ}}$ and frequencies $\nu_{\pm}$.
By coherently exciting both $\ket{\pm 1}$ levels, a spin-1 qutrit superposition is prepared that enhances the phase accumulation and sensitivity to the oscillating DM field compared to a conventional qubit configuration.}
    \label{fig:dm_schematic}
\end{figure}

To achieve maximum sensitivity, the system is initialized in a state of maximal spin-projection variance, corresponding to a balanced superposition of the extremal eigenstates of $\hat S_z$. For a qubit restricted to the $\{\ket{m=0},\ket{m=+1}\}$ subspace, the superposition $(\ket{0}\pm\ket{+1})/\sqrt2$ yields $\langle \Delta\hat S_z^2\rangle_{\rm qubit}=1/4$.
For qutrits, employing the full spin-1 space preparing a superposition of the states with maximally separated magnetic quantum numbers $(\ket{-1}\pm\ket{+1})/\sqrt2$ yields $\langle \Delta\hat S_z^2\rangle_{\rm qutrit}=1$.
In Fig.~\ref{fig:dm_schematic}, we illustrate the difference between qubit and qutrit systems targeting ultralight wave-like DM detection.
The enhancement corresponds to a fourfold increase in QFI
\begin{equation} \label{eq:QFI}
    \frac{{\rm QFI}_{\rm qutrit}}{{\rm QFI}_{\rm qubit}} = \frac{4\langle\Delta \hat{S}_z^2\rangle_{\rm qutrit}}{4 \langle\Delta \hat{S}_z^2\rangle_{\rm qubit}} = 4~.
\end{equation}
Therefore, a qutrit sensor, neglecting decoherence effects, can achieve a universal factor of two improvement in amplitude sensitivity (i.e. factor four gain in power) over a qubit, regardless of the specific control sequence.

Saturating the QFI Cramer-Rao bound requires not only optimal state preparation but also optimal measurements.
This measurement is defined by the symmetric logarithmic derivative (SLD) operator $\hat{L}_\theta$ associated with the parameter $\theta$~\cite{Liu:2019xfr}, 
whose eigenbasis specifies the projective measurement that achieves equality in the bound.  
For a conventional qubit system  this corresponds to projection along the spin axis orthogonal to the initial $\pi/2$ rotation.  
We derive the explicit form of $\hat L_\theta$ and its eigenstates for a qutrit system in the Supplemental Material.
Our analysis   generally applies to any spin-1 qutrit system where an external field couples linearly to $\hat S_z$. 

This enhancement arises solely from the spin-projection variance  and thus generalizes to other spin-$J$ systems with quantum number separation $\Delta m = 2 J$. The result establishes a universal scaling of sensitivity with Hilbert space dimensionality, independent of pulse sequence or physical platform.

\prlsection{Filtering function enhancement}{.}
We now demonstrate that the same universal enhancement can be understood from the filter function formalism, which describes how control pulse sequences impose external field fluctuations into sensor decoherence by filtering the spectral density of a noise source~\cite{Barry:2019sdg}.  
Consider a superposition between two spin eigenstates separated by $\Delta m$.  
For a qubit state $(\ket{0}+\ket{+1})/\sqrt2$ one has $\Delta m=1$, while for a qutrit state $(\ket{-1}+\ket{+1})/\sqrt2$ one has $\Delta m=2$.  

After an interrogation time $t$, each state $\ket{m_i}$ accumulates a phase $\theta_i = m_i \theta$, where $ \theta(t) = \gamma_e \int_0^t f(t, t') \mathcal{B}(t') \mathrm{d}t'$. Here, as before, $\mathcal{B}(t')$ denotes the effective and possibly novel magnetic or pseudo-magnetic field and $f(t,t')$ is the control function determined by the pulse sequence.

Averaging over stochastic fluctuations of $\mathcal{B}(t')$ gives the coherence function, which can be obtained directly from the density matrix formalism (see Supplemental Material for details) where $\rho_{m_1, m_2}$ is the density matrix element between spin eigenstates $\ket{m_1}$ and $\ket{m_2}$
\begin{align}
W(t)= \dfrac{|\langle\rho_{m_1m_2}(t)\rangle|}{|\langle\rho_{m_1m_2}(0)\rangle|}
  = \Big| \Big\langle e^{-i\Delta m \theta} \Big\rangle \Big|
  = e^{-\chi(t)}.
\end{align}
Here, considering Gaussian fluctuations with $\langle \mathcal{B}(t')\rangle=0$, the decoherence exponent is~\cite{Paz-Silva:2014gkk,Szankowski:2017msc}
\begin{align}
\chi(t)= \int_0^\infty \frac{d\omega}{\pi} 
\mathcal{S}_{\omega} \frac{F(\omega t)}{\omega^2},
\end{align}
with $\mathcal{S}_{\omega}$ denoting the field's spectral density power spectrum that is proportional to the autocorrelation of $\mathcal{B}$ 
\begin{align}
    \mathcal{S}_\omega \propto |\Delta m|^2 \int d\tau \average{\mathcal{B}(t)\mathcal{B}(t+\tau)} e^{-i \omega \tau}~,
\end{align}
and $F(\omega t)=(\omega^2/2)|\tilde f(t;\omega)|^2$ the filter function of the control sequence with $\tilde{f}(t;\omega)$ being the Fourier transform of the control function.  

Since $\mathcal{S}_\omega \propto |\Delta m|^2$, the signal power and thus the induced phase variance scale as $|\Delta m|^2$.
Therefore, a qutrit superposition with $\Delta m=2$ yields a factor of four greater response than a qubit with $\Delta m=1$ independently of the control sequence, corresponding to a factor of two gain in amplitude sensitivity.  
This reproduces our QFI result and can be interpreted as reflecting the geometric origin of the enhancement with the greater separation of eigenstates in the higher dimensional Hilbert space increasing the distinguishability of phase-shifted states.
 
In addition to the enhancement from the increased spin-precession rate, measurements exploiting the full spin-1 manifold are inherently differential. That is, common-mode noise sources that perturb both $\ket{ +1}$ and $\ket{ -1}$ levels in the same direction, such as temperature drifts, axial electric fields, or transverse magnetic fields, effectively cancel for qutrits~\cite{Acosta2010,Toyli2013,Kucsko2013,Barry:2019sdg}. Such effects are system dependent.
This differential character improves robustness and further enhances the sensitivity of certain qutrit systems.

In practice, the actual realized gain depends on the dominant decoherence mechanism and various noise contributions beyond readout.
If the environment is dominated by magnetic spin bath noise, the twice-enhanced magnetic response of the qutrit system shortens the dephasing time $T_2^*$ leading to broadening the linewidth and partially offsetting the sensitivity gain~\cite{Barry:2019sdg,Fang2013,Mamin2014,Kucsko2016,Bauch2018}.  
Conversely, when non-magnetic common-mode noise sources such as temperature or electric field fluctuations prevail for the specific qutrit system, qutrit sensing can outperform qubit setups even further, as common-mode rejection suppresses such noise, and the effective $T_2^*$ could actually be extended.

\prlsection{Nitrogen-vacancy center metrology}{.}
As a concrete realization of this general principle, we consider the spin-1 system of    the negatively-charged NV center in diamond ($\mathrm{NV}^-$) that is particularly well suited for quantum sensing applications and  naturally provides a controllable three level qutrit system with well developed quantum control protocols~\cite{Degen:2016pxo}. In this state, two electrons form a spin triplet ($S=1$) with sublevels $\ket{m=0}$ and $\ket{m=\pm1}$ corresponding to spin projections along the crystal's $z$ axis.  
The $\ket{m=\pm1}$ states define the double-quantum (qutrit) space.

\begin{figure}[t]
    \centering
    \includegraphics[width=0.99\linewidth]{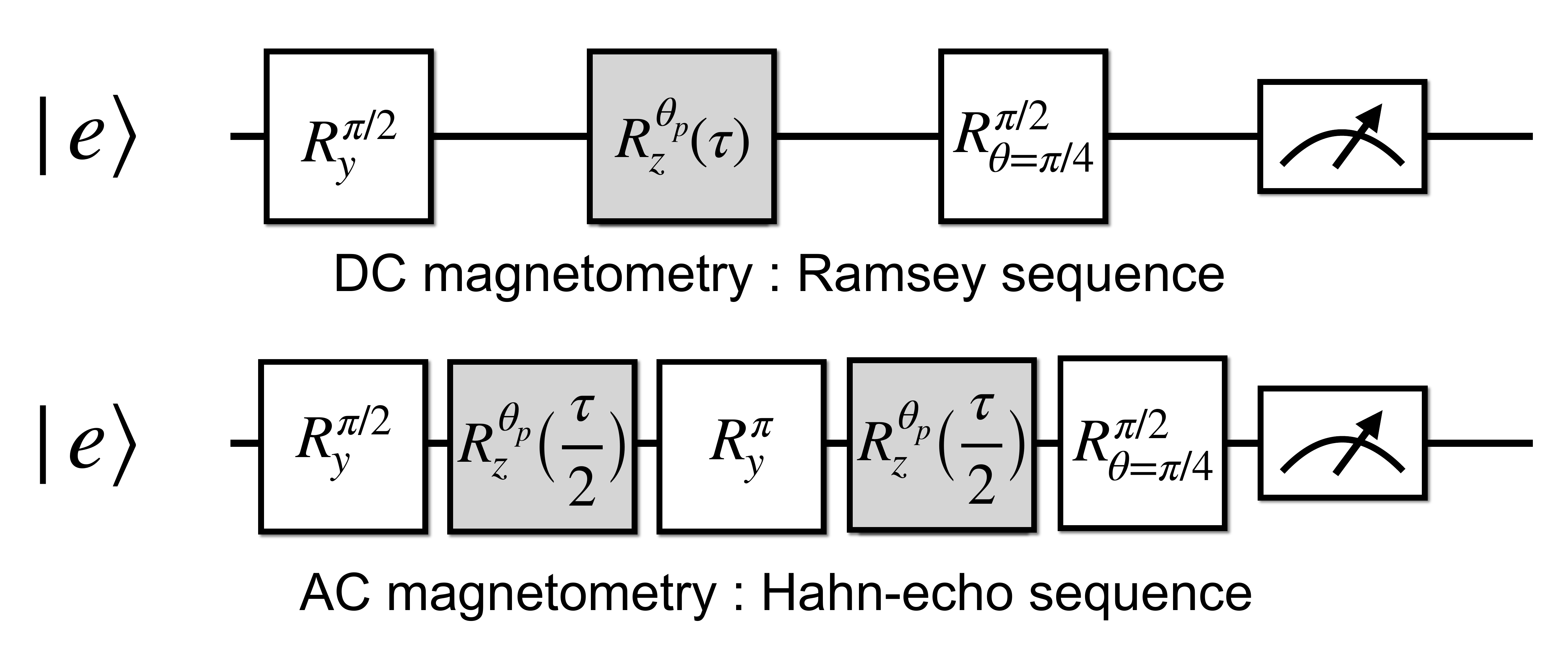} 
    \caption{ \label{fig:sequence}
Pulse sequence protocols for DC and AC magnetometry using NV center electron spins as qutrits.
\textit{Top}: For DC magnetometry, a Ramsey sequence is applied. A $\pi/2$ rotation about the $y$-axis ($R_y^{\pi/2}$) prepares a coherent superposition, which freely evolves under $R_z^{\theta_p}(\tau)$ to accumulate a magnetic field-dependent phase, followed by a second $\pi/2$ pulse with phase $\theta = \pi/4$ and optical readout.
\textit{Bottom}: In AC magnetometry a Hahn-echo sequence is applied and suppresses low-frequency noise. After the initial $\pi/2$ pulse, free evolution for $\tau/2$ is interrupted by a refocusing $\pi$ pulse and followed by another $\tau/2$ evolution and final $\pi/2$ pulse.
The acquired phase encodes the oscillating magnetic-field component resonant with the sequence.
}
\end{figure}

Double-quantum operation in NV ensembles has been shown to extend the spin-dephasing time $T_2^*$ by up to a factor of eight compared to single-quantum measurements~\cite{Bauch:2018unc}, confirming that the expected qutrit advantage in coherence time can be realized under realistic conditions.
Recent experiments achieved $T_2^*=28.6~\mu\mathrm{s}$ ($T_2=324~\mu\mathrm{s}$) for Ramsey (Hahn-echo) qutrit protocol sequences~\cite{Barry:2023hon}, which apply spin-bath driving as well for further coherence time improvements~\cite{Bauch:2018unc}. These are significant improvements compared to $T_2^*=1~\mu\mathrm{s}$ ($T_2\sim  100~\mu\mathrm{s}$) for the corresponding qubit configurations used in previous work~\cite{Chigusa:2023roq,Wolf:2014ihu}. We illustrate the two qutrit protocols in Fig.~\ref{fig:sequence}.
The qutrit parameters above are adopted in our subsequent sensitivity analysis.

Denoting the total electron spin operator as $\vec S_e$, the interaction Hamiltonian is given by~\cite{Barry:2019sdg} 
\begin{align}
    H = 2\pi D S_{e,z}^2 + \gamma_e B_0 S_{e,z} + \gamma_e \vec B\cdot \vec S_e,
    \label{Hamiltonian}
\end{align} 
where $D=2.87 {\rm GHz}$ is the zero-field splitting parameter, $B_0$ is an applied bias magnetic field along the $z$-axis, $\vec B = (B_x, B_y, B_z)^t$ is an external magnetic field to be sensed. The energy difference between the ground state $\left|0\right>$ and the excited states $\left|\pm\right>$ is given by $\omega_\pm = 2\pi D \pm \gamma_e B_0$. For a time varying and weak magnetic field $\vec{B}(t)$, the last term in Eq.~\eqref{Hamiltonian} can be taken as perturbation, see the Supplemental Material.

We find that the Ramsey-type sequence that prepares the initial superposition and performs a final $\pi/2$ rotation at $45^\circ$ relative to the first pulse gives the optimal sensitivity.
This measurement basis is equivalent to the eigenbasis of the SLD. Hence, this explicitly demonstrates how the QFI qutrit bound can be saturated with realistic measurements. The details of the derivation are provided in the Supplemental Material.

Using ensemble readout, the dominant fundamental noise source is spin-projection noise.  
Assuming $N$ centers and an observation time $t_{\mathrm{obs}}$, the sensitivity scales as $\Delta B_{\mathrm{sp}} \propto e^{\tau/T_2^*}/(\gamma_e\sqrt{N \tau t_{\mathrm{min}}}|\Delta m|)$, with $N$ the number of NV centers in the ensemble and $t_{\mathrm{min}}$ the minimum of the DM coherence time and the total observation time (see Supplemental Material for details).
For the parameters above, this corresponds to $\Delta B_{\mathrm{sp}} \sim 1~\mathrm{fT}$ for a $1$-second observation time with $N=10^{12}$ NVs, demonstrating that the higher dimensionality and the longer $T_2^*$ attainable in qutrit operation directly improve the sensitivity to weak fields compared to previous work~\cite{Chigusa:2023roq}.

\begin{figure*}[t]
    \centering
    \includegraphics[width=0.49\linewidth]{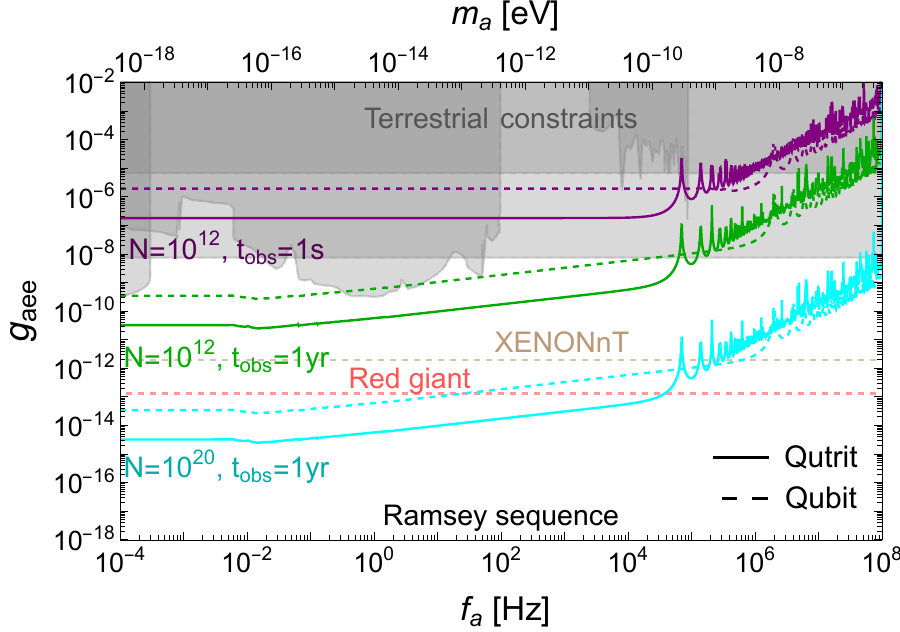}
    \includegraphics[width=0.49\linewidth]{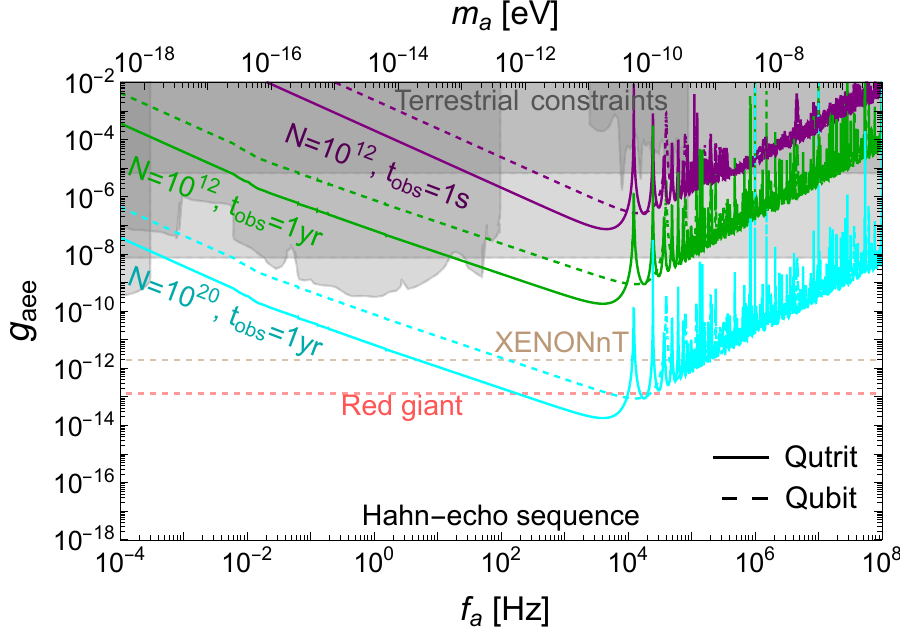}
    \caption{\textit{Left panel}: The projected $95\%$ confidence level sensitivity to the axion-electron coupling ($g_{aee}$) for NV qutrit (solid line) compared with qubit results of Ref.~\cite{Chigusa:2023roq} (dashed line) using DC magnetometry Ramsey sequences. The DC magnetometry has a broadband sensitivity for the low frequency region of DM.  For frequencies smaller than the Nyquist frequency $f_a\leq 1/\tau$, owing to improved signal strength and the longer dephasing time of qutrits, they have about an order of magnitude improvement. \textit{Right panel}: As the left panel, but for AC magnetometry with the Hahn-echo sequence. AC magnetometry is highly sensitive to the higher frequency  range of $f_a\sim 1/\tau$, while quickly losing sensitivity for both lower and higher frequencies. For the frequency $f_a\leq 1/\tau$, the qutrit shows an improvement of around an order of magnitude. For each system and sequence we considered the value $\tau=T_2^{(*)}/2$, with the dephasing time (transverse relaxation time) $T_2^{(*)}$   given in main text. The terrestrial laboratory limits include comagnetometers~\cite{Bloch:2019lcy}, torsion pendulum experiments ~\cite{Terrano:2015sna,Terrano:2019clh} and fermionic axion interferometer~\cite{Crescini:2023zyl}, with underground detector limits from XENONnT~\cite{XENON:2022ltv} considering solar axions and red giants~\cite{Capozzi:2020cbu} also  shown.}
\label{fig:ramsey_sensitivities}
\end{figure*}

\prlsection{Dark matter detection}{.} 
Ultralight axions and ALPs constitute well motivated DM candidates, arising naturally from solutions to the strong CP problem and in string theories~\cite{peccei1977cp,Svrcek:2006yi,Arvanitaki:2009fg}. Their relic abundance can be produced through various cosmological mechanisms~\cite{Preskill:1982cy,Abbott:1982af,Dine:1982ah}, and in the local Galactic halo they behave as a non-relativistic coherently oscillating classical field, $a(t)$, with a frequency set by their mass, $\omega \simeq m_a$. Depending on the underlying theory and coupling to Standard Model, a variety of axion and ALP signatures can arise and are targeted experimentally~\cite{Raffelt:1990yz,Graham:2015ouw,Safronova:2017xyt,Adams:2022pbo}. 

We focus on the axion-electron coupling that is generically predicted in a broad range of scenarios.
The interaction Lagrangian  is given by
\begin{equation}
    \mathcal{L}\supset g_{aee}\frac{\partial_\mu a}{2m_e}\bar{\psi}_e\gamma^\mu \gamma_5 \psi_e.
\end{equation}
where $g_{aee}$ is the dimensionless coupling constant and $a$ is the axion field and $\psi_e$ denotes the electron field. In the non-relativistic limit, these interactions induce an effective magnetic field that acts on the electron spin in the NV center. The effective Hamiltonian is
\begin{equation} 
    H_{\rm eff} =\frac{g_{aee}}{m_e}\vec{\nabla}a\cdot \vec{S}_e = \gamma_e \vec B_{\rm eff} \cdot \vec{S}_e.
\end{equation} 
For a coherently oscillating Galactic DM ALP field $a(t)=a_0\cos(m_a t+\delta)=(\sqrt{2\rho_{\rm DM}}/m_a)\cos(m_a t+\delta)$ with local density $\rho_{\rm DM}\simeq0.4~{\rm GeV cm^{-3}}$ and $\delta$ random phase, the NV center system will experience an effective pseudomagnetic field
$\vec B_{\rm eff} 
        \simeq (   g_{aee}/\gamma_e m_e)  \sqrt{2\rho_{\rm DM}}v_{\rm DM}\cos(m_a t +\delta),
$ where $v_{\rm DM} \sim 10^{-3}$ is the characteristic cold velocity of the DM. Here, we model the axion DM field as a simple monochromatic plane wave temporarily neglecting its inherent stochastic nature.
This oscillating field drives spin precession in the NV-center qutrit, producing the measurable phase shift discussed above. 
For our analysis we adopt the improved  method following Ref.~\cite{Dror:2022xpi,Chigusa:2024psk} that imposes a cut-off on signal correlations beyond the DM coherence time to account for stochastic nature.

In Fig.~\ref{fig:ramsey_sensitivities} we show the projected $95\%$ confidence level sensitivity to the axion-electron coupling for Ramsey and Hahn-echo sequences using NV qubits and qutrits, assuming the spin projection noise-limited regime (see Supplemental Material for details). The projected qutrit reach improves upon previously published qubit one by up to an order of magnitude~\cite{Chigusa:2023roq}. This enhancement arises from both the increased QFI of the multilevel qutrit system and the extended coherence time, the latter benefiting from the qutrit configuration as well as from spin-bath driving~\cite{Barry:2023hon}. For the Hahn-echo sequence, while the coherence time improvement is smaller, the net sensitivity gain is comparable (see Supplemental Material).

The parameter choice in Fig.~\ref{fig:ramsey_sensitivities}, such as the number of NV centers, follows previous work for direct comparison~\cite{Chigusa:2023roq}. Several experimental considerations are worth noting. Although ensembles containing $10^{20}$ NVs are not yet available, higher yields recently achieved for NV creation improving from $0.68\%$ in Ref.~\cite{Barry:2023hon} to $25.8\%$ in Ref.~\cite{Balasubramanian:2022} together with the use of thicker or multiple samples can significantly increase the effective number of sensors. Further, our results demonstrate that qutrit-based experiments with available sample sizes could already improve on existing  terrestrial constraints from comagnetometers~\cite{Bloch:2019lcy}. 

In addition, for NV readout the limiting noise source is photon shot noise rather than spin-projection noise, which reduces absolute sensitivities by approximately an order of magnitude (see Supplemental Material). However, the relative advantage of qutrits over qubits remains unchanged, highlighting our results as a significant step forward in the sensitivity of NV-based searches for fundamental interactions. Further improvements could arise from employing squeezed states, for which there is recent progress~\cite{Wu:2025jfo}, and that could potentially enhance sensitivities for qutrits compared to qubits~\cite{Gassab:2024izv}. 

\prlsection{Conclusions}{.}
We have demonstrated that harnessing higher dimensional quantum states can provide a universal, sequence-independent enhancement in sensing precision. To our knowledge, this is the first demonstration that multilevel quantum states can enhance sensitivity to fundamental interactions, establishing a new framework for quantum sensing beyond qubits. For spin-1 qutrits this yields a four-fold increase in QFI and a two-fold gain in projection noise-limited sensitivity. Applied to ultralight DM searches with NV centers in diamond, our realizable protocols improve the axion-electron coupling reach by up to an order of magnitude compared to qubit schemes. This enhancement arises not only from the intrinsic QFI scaling but also from extended coherence times enabled by the common-mode noise suppression of the NV-center qutrit system. Even further advantages could emerge with the implementation of spin squeezing in multilevel systems.

Since the QFI enhancement follows from the algebra of spin operators, rather than any NV specific property, the same multilevel sensing principle applies broadly to other systems such as superconducting, trapped-ion  and neutral-atom qutrits or higher dimensional qudits. This establishes a general quantum metrological scaling for multilevel systems and positions them as a new frontier for precision tests of fundamental physics.

~\newline
\textbf{Acknowledgments.---}
The authors thank Kazunori Nakayama for participating in the early stages of this work. 
X.L.M. thanks Tengyu Ai for helpful discussion.
The work was supported by the World Premier International Research Center Initiative (WPI), MEXT, Japan. V.T. acknowledges support from JSPS KAKENHI grant No. 23K13109. The work of N.M., E.D.H. was partially supported by MEXT Q-LEAP No. JPMXS0118067395.

\setlength{\bibsep}{3pt} 
\bibliography{references}

\clearpage
\newpage  
\appendix
\onecolumngrid
\renewcommand{\theequation}{S.\arabic{equation}}
\setcounter{equation}{0}

\centerline{\large {Supplemental Material for}}
\medskip

{\centerline{\large \bf{Beyond Qubits: Multilevel Quantum Sensing for Dark Matter}}}
\medskip
{\centerline{Xiaolin Ma, Volodymyr Takhistov, Norikazu Mizuochi, Ernst David Herbschleb}}
\bigskip
\bigskip
  
\vspace{1em} 
\counterwithin{figure}{section}
\counterwithin{table}{section}

In this Supplemental Material we provide additional details on the formalism of NV-center metrology with qutrits, the saturation of the quantum Fisher information bound, the computation of signal and noise power spectral densities  and the scaling of the axion DM sensitivity.

\section{A. Formalism for Nitrogen-Vacancy Metrology with Qutrits}

Here, we describe the quantum-mechanical control of the NV center electron spin qutrit using microwave (MW) pulses.
The system dynamics are governed by the time-dependent Schr\"{o}dinger equation 
\begin{equation}
    i\hbar \frac{d}{dt}\ket{\psi(t)} = H(t)\ket{\psi(t)},
\end{equation}
with the total Hamiltonian 
\begin{align}
    H(t) = H_0 + H_1(t),
\end{align}
where the static Hamiltonian $H_0$ and the time-dependent interaction   $H_1(t)$ are
\begin{align}
    H_0 &= 2\pi D \hat{S}_{e,z}^2 + \gamma_e B_0 \hat{S}_{e,z}, \\
    H_1(t) &= \gamma_e \vec{B}_{\rm eff}(t) \cdot \vec{\hat{S}}_e.
\end{align}
Here, $D = 2.87~\text{GHz}$ is the zero-field splitting with $2\pi D \simeq 11.9~\mu\text{eV}$,  $B_0$ is an applied bias magnetic field along the NV symmetry axis  taken as $z$-axis, 
and $\vec{B}_{\mathrm{eff}}(t)$ represents the external, time-dependent magnetic field to be measured.

In the spin-$1$ basis $\{\ket{+1}, \ket{0}, \ket{-1}\}$  the general magnetic interaction term can be written as 
\begin{align}
    \gamma_e \vec{B} \cdot \vec{\hat{S}}_e = \gamma_e \begin{pmatrix}
        B_z & \dfrac{B_-}{\sqrt 2}   & 0 \\
        \dfrac{B_+}{\sqrt 2}   & 0 &  \dfrac{B_-}{\sqrt 2}   \\
        0 & \dfrac{B_+}{\sqrt 2}   & -B_z
    \end{pmatrix},
\end{align}
where $B_\pm = B_x \pm iB_y$.  
The interaction term $H_1(t)$ is treated as a perturbation, since for the fields of interest we consider $\vec{B}_{\mathrm{eff}}$ is assumed to be weak.

To solve the dynamics  we consider the interaction picture, defining
 $\ket{\psi_I(t)} = e^{iH_0 t} \ket{\psi_S(t)}$. Dephasing is neglected in this evolution and 
its effect can be included subsequently as an additional exponential factor $e^{-\tau/T_2^*}$ in the sensitivity calculation.
Hence, the evolution  becomes
\begin{align}
    i \frac{\partial}{\partial t} \ket{\psi_I(t)} = H_{I}(t) \ket{\psi_I(t)}, \quad
    H_{I}(t) = e^{i H_0 t} H_1(t) e^{-i H_0 t}.
\end{align}
Explicitly, the interaction Hamiltonian in the spin-1 Hilbert space takes the form
\begin{align}
H_I(t) = \gamma_e \begin{pmatrix}
 B_z(t)    & \dfrac{B_-(t)}{\sqrt 2}e^{i\omega_+ t} & 0 \\
        \dfrac{B_+(t)}{\sqrt 2} e^{-i\omega_+ t} & 0 &  \dfrac{B_-(t)}{\sqrt 2} e^{-i\omega_- t}  \\
        0   &  \dfrac{B_+(t)}{\sqrt 2} e^{i\omega_- t} & -B_z(t) \\
    \end{pmatrix},
\end{align}
where $\omega_\pm = 2\pi D \pm \gamma_e B_0$ are the transition frequencies between the $\ket{0}$ and $\ket{\pm1}$ states, respectively.

To manipulate the qutrit state, we employ Hamiltonian engineering using balanced double-driving MW pulses that simultaneously drive both the $\ket{0} \leftrightarrow \ket{+1}$ and $\ket{0} \leftrightarrow \ket{-1}$ transitions with equal amplitude~\cite{Zhou:2023xnx}. 
We consider a driving field polarized along the $y$ direction,
\begin{equation}
    \vec{B}_{\mathrm{drive}}(t)
    = \hat{y}  B_1  \left[\cos(\omega_+ t) + \cos(\omega_- t)\right],
\end{equation}
which temporarily replaces $\vec{B}_{\mathrm{eff}}$ in $H_1$.

Applying the rotating wave approximation (RWA) and neglecting fast-oscillating terms, the interaction Hamiltonian simplifies to
\begin{align}
    H_I \simeq \dfrac{\gamma_e B_1}{2\sqrt{2}} \begin{pmatrix}
        0 & -i & 0 \\
        i & 0 & -i \\
        0 & i & 0
    \end{pmatrix} = \dfrac{\gamma_e B_1}{2} \hat{S}_y.
\end{align}
The corresponding evolution operator represents a rotation about the $y$ axis: $U(t) = e^{-i H_I t} = \exp(-i (\gamma_e B_1 t/2) \hat{S}_y)$. Its  matrix form is
\begin{align}
e^{-i H_I t} &= I - i \hat{S}_y \sin\left(\dfrac{\gamma_e B_1t}{2}\right) + (\hat{S}_y)^2 \left[\cos\left(\dfrac{\gamma_e B_1t}{2}\right)-1\right] \\
&= \begin{pmatrix}
     \cos^2\left(\dfrac{\phi_1}{4}\right) & -\frac{1}{\sqrt{2}}\sin\left(\dfrac{\phi_1}{2}\right) & \sin^2\left(\dfrac{\phi_1}{4}\right)  \\
       \dfrac{1}{\sqrt{2}}\sin\left(\dfrac{\phi_1}{2}\right) & \cos\left(\dfrac{\phi_1}{2}\right) & - \dfrac{1}{\sqrt{2}}\sin\left(\dfrac{\phi_1}{2}\right) \\
      \sin^2\left(\dfrac{\phi_1}{4}\right)  &  \dfrac{1}{\sqrt{2}}\sin\left(\dfrac{\phi_1}{2}\right) & \cos^2\left(\dfrac{\phi_1}{4}\right) \end{pmatrix},
\end{align}
where $\phi_1 = \gamma_e B_1 t$. We have verified that restricting this matrix to the top-left $2\times 2$ block corresponding to $\{\ket{+1}, \ket{0}\}$ subspace reproduces the standard qubit results~\cite{Chigusa:2023roq}.

More generally, for a pulse in the $x$-$y$ plane at an angle $\theta$ relative to the $y$-axis, the gate operator is \begin{equation}
    R_\theta(\phi)
    = \exp \left[-i(\hat{S}_y\cos\theta + \hat{S}_x\sin\theta)\phi\right],
\end{equation}
where $\phi$ is the pulse area.  By controlling the pulse duration $t$, amplitude $B_1$, and phase $\theta$  continuous rotations of the qutrit can be implemented within the full spin-1 Hilbert space.

The dynamics described  enable the design of pulse sequences for NV-center qutrit magnetometry, directly analogous to the qubit case~\cite{Chigusa:2024psk,Chigusa:2023roq}.  
Standard sequences such as Ramsey~\cite{ramsey1950molecular}, Hahn-echo~\cite{Hahn1950}, and CPMG~\cite{hung2004application,casabianca2014chirped} can be implemented.  
A typical sensing protocol prepares a superposition state, allows it to precess and accumulate a phase from an external field  and finally maps this phase onto a population difference for optical readout.

As the qutrit analog of conventional DC magnetometry, we employ a Ramsey-type sequence~\cite{Barry:2019sdg,Chigusa:2023roq,Chigusa:2024psk}:
\begin{itemize}
    \item[(i)] The system is initialized in the $\ket{0}$ state.  
    A double-driving $\pi/2$ pulse, as described, simultaneously excites the $\ket{\pm1}$ states, creating the superposition $(\ket{+1}-\ket{-1})/\sqrt{2}$.
    
    \item[(ii)] The state freely precesses for a time $\tau$, during which a relative phase accumulates due to the external field.  
    In practice, $\tau$ is chosen to be comparable to the spin dephasing time $T_2^* \sim \mathcal{O}(10) \mu\mathrm{s}$~\cite{Barry:2023hon}.  
    For simplicity, we take the optimal precession time $\tau = T_2^*/2$ throughout this work~\cite{Herbschleb:2019zlq}. The dephasing time $T_2^*$ depends on the NV-center concentration.  
    For clarity, this dependence is neglected here, as the total number of NV centers can be varied through both concentration and sample volume. 
    
    \item[(iii)] A second pulse is applied with its MW field tilted by an angle $\theta$ relative to the initial $\pi/2$ pulse  followed by a finite pulse duration.  
    
    \item[(iv)] The final state populations are measured via fluorescence in the NV-center readout.
\end{itemize}
Compared with the standard qubit case, the observable here is $\hat{S}_z^2$ rather than $\hat{S}_z$, reflecting the three level structure and the fact that optical readout cannot distinguish between the $\ket{+1}$ and $\ket{-1}$ states~\cite{Barry:2019sdg}.  
This particular sequence is also found to be optimal according to the QFI analysis presented in the following section.

\section{B. Qutrit Saturation of Quantum Fisher Information Bound}

As discussed in the main text, the sensitivity enhancement achieved by employing a qutrit compared to a qubit can also be understood from the perspective of QFI.  
The QFI $F_q$  sets the ultimate precision limit for estimating a parameter $\theta$ encoded in a quantum state, as formalized by the quantum Cramer-Rao bound $(\Delta \theta)^2 \geq 1/F_q $. A larger QFI corresponds to a smaller attainable variance in $\theta$ and thus greater sensitivity~\cite{Cowan:2010js,Helstrom1969QuantumDA,holevo2011probabilistic}.

Consider an initial state $\ket{\psi(0)}$ that evolves under effect of a field that imparts a phase $\theta$ through the spin operator $\hat{S}_z$.
Starting from the perturbed state and Eq.~\eqref{eq:perturbe_state} 
\begin{align}
    |\psi\rangle=\exp(i \hat{S}_z \theta) |\psi(0)\rangle.
\end{align}
The accumulated phase $\theta$ depends on the effective magnetic field, in our case originating from axion DM, $\mathcal{B}(t)$, and the control function $y(t)$ of the pulse sequence, $\theta=\int \mathcal{B}(t) y(t) dt$.
Here,  $y(t)=\pm1$ represents the switching function depending on the sequence protocols and physically implemented by $\pi$ pulses. For a pure state undergoing unitary evolution, the QFI is given by~\cite{Liu:2019xfr} 
\begin{align}
    F_q(\theta)=4(\langle \partial_\theta \psi|\partial_\theta \psi \rangle -|\langle \partial_\theta \psi|\psi \rangle |^2) =4\langle\Delta \hat{S}_z^2\rangle,
\end{align}
where $\langle\Delta \hat{S}_z^2\rangle = \langle\hat{S}_z^2\rangle - \langle\hat{S}_z\rangle^2$ is the variance of the spin operator $\hat{S}_z$ in the initial state $|\psi(0)\rangle$. 
Maximizing the QFI therefore requires maximizing $\langle \Delta \hat{S}_z^2 \rangle$, which is achieved when the initial state is a maximal superposition of the two magnetic sublevels with the largest possible $\Delta m$, that is
$\left(\ket{m_1}  \pm  \ket{m_2}\right)/\sqrt{2}$ with $\Delta m = |m_1 - m_2| = 2J$ for a spin-$J$ system.

For a standard qubit utilizing the $\{ \ket{m=0}, \ket{m=+1} \}$ subspace, we prepare the state $(\ket{0}+\ket{+1})/\sqrt{2}$. 
This yields $\langle \hat{S}_z \rangle = 1/2$ and $\langle \hat{S}_z^2 \rangle = 1/2$, giving a variance
$\langle  \Delta \hat{S}_z^2 \rangle_{\mathrm{qubit}} = 1/4$.  
For a qutrit, the full spin-1 space can be exploited by preparing a superposition of the states with the largest magnetic quantum number separation $\ket{\psi_{\mathrm{qutrit}}} = (\ket{-1}+\ket{+1})/\sqrt{2}$.
For this state, $\langle \hat{S}_z \rangle = 0$ and $\langle \hat{S}_z^2 \rangle = 1$, 
resulting in $\langle  \Delta \hat{S}_z^2 \rangle_{\mathrm{qutrit}} = 1$.  
The ratio of the corresponding QFIs is
\begin{align}
    \frac{{\rm QFI}_{\mathrm{qutrit}}}{{\rm QFI}_{\mathrm{qubit}}}
    = \frac{\langle \Delta \hat{S}_z^2 \rangle_{\mathrm{qutrit}}}
           {\langle \Delta \hat{S}_z^2 \rangle_{\mathrm{qubit}}}
    = 4,
\end{align}
indicating that, neglecting decoherence effects for simplicity, independent of the specific pulse sequence  a qutrit sensor can in principle achieve enhancement of twice the amplitude sensitivity and four times the power of a qubit sensor.

Saturating the QFI bound requires in addition to optimal state preparation also optimal measurements.  
This can be analyzed considering SLD operator associated with the parameter $\theta$ is defined as~\cite{Liu:2019xfr}
\begin{align}
  \hat{L}_\theta=2(  |\psi\rangle \langle \partial_\theta \psi|-|\partial_\theta \psi\rangle \langle \psi |).
\end{align}
The optimal measurement projects onto the eigenbasis $\{\ket{\phi_i}\}$ of $\hat{L}_\theta$, with projectors $\hat{\Pi}_i=\ket{\phi_i}  \bra{\phi_i}$ giving probabilities ${\rm Tr}[\hat{\Pi}_i \rho]$.  
For a standard qubit, the optimal measurement corresponds to projection along the spin axis orthogonal to the initial $\pi/2$ (Hadamard) rotation~\cite{Chigusa:2023roq,Chigusa:2024psk}.

For the qutrit state $\ket{\psi_{\mathrm{qutrit}}}$, the SLD operator can be obtained as
\begin{align}
  \hat{L}_\theta=-2i\begin{pmatrix}
    0 & 0 & e^{2i \theta_p} \\
    0 & 0 & 0 \\
    e^{-2i \theta_p} &0 & 0
    \end{pmatrix}
    \simeq-2i\begin{pmatrix}
    0 & 0 & 1 \\
    0 & 0 & 0 \\
    1 &0 & 0
    \end{pmatrix}.
\end{align}
where $\theta_p$ is the accumulated phase during free precession.  
The corresponding optimal measurement operator is
$\hat{\Pi} = \sum_i \ket{\phi_i} \bra{\phi_i}$, and the optimal measurement that saturates the QFI bound gives $\Tr[\hat{\Pi}_i \rho]$.
This can be implemented experimentally by applying a final rotation $\hat{R}$ prior to fluorescence readout in the $\hat{S}_z^2$ basis, such that 
$\hat{R}^\dagger \hat{S}_z^2 \hat{R}$ matches the SLD eigenbasis.  
This condition is met for a $\pi/2$ pulse tilted by $\theta = \pi/4$ relative to the initial pulse along the $y$ axis.  
Thus, from the QFI perspective, the designed sequence achieves the optimal measurement that saturates the QFI bound for detecting weak fields such as those from DM.

As an explicit illustration, consider the final state after the qutrit Ramsey sequence 
\begin{align}
    \left|\psi_I(\tau)\right> &= U_3^{x_3,~\theta}  U_2^{\rm free}  U_1^{\pi/2} \ket{0},
    \label{psi_fin_DC}
\end{align}
where $U_1$ and $U_3$ denote control pulses 
 and 
\begin{equation}
    U_2^{\mathrm{free}}
    = \hat{T} \left[e^{-i \int_0^\tau H_I^{\mathrm{eff}}dt}\right]
    = {\rm diag}(e^{i\theta_p},1,e^{-i\theta_p})
\end{equation} 
represents the phase accumulated from the external field  with 
$\theta_p = \gamma_e  \int  B_z(t')dt'$.  
At the end of the sequence  the ensemble fluorescence is measured.  
Since the emitted photons from $\ket{\pm1}$ are indistinguishable, the observable is proportional to $\langle \hat{S}_z^2 \rangle$.  
The resulting qutrit signal is
\begin{align}
    S = 
    \bra{\psi_I(\tau)} \hat{S}_z^2 \ket{\psi_I(\tau)}
    \simeq
    \sin(2\theta)\sin^2(\phi) \theta_p + {\rm const},
    \label{S_fin_DC}
\end{align}
where the small signal limit $\theta_p \ll 1$ is taken. The constant term, independent of the magnetic field, can be subtracted during analysis.  
The signal is maximized for $\theta=\pi/4$ and $x_3=\pi/2$, corresponding to a $\pi/2$ pulse about an axis rotated by $45^\circ$ from the initial pulse.  
This differs from the qubit case, where the final pulse is orthogonal to the initial one.  
Remarkably, this same pulse configuration also satisfies the condition for saturating the QFI bound, confirming that the Ramsey-type qutrit sequence constitutes the optimal protocol for sensing weak fields.

\section{C. Computation of Signal and Noise Power Spectral Density}
\label{sec:refined}

\subsection{C.1: Signal power}

In this section, we provide a detailed derivation of the signal and background power spectral densities (PSDs) for a general pulse sequence, following qubit analysis of Ref.~\cite{Chigusa:2024psk}.  
After a sequence of duration $\tau$, the density matrix of a single NV-center qutrit $\ell$ in the $j$-th measurement is $  \rho_{j\ell} = \ket{\psi(\tau)} \bra{\psi(\tau)}$.  To remain general, we define a measurement operator $\hat{d}_{j\ell}$.  
For an $N$  ensemble of   NV centers, the total observable is  $\hat{D}_j = \sum_{\ell=1}^{N} \hat{d}_{j\ell}$  
and the ensemble density matrix can be approximated as 
$\rho_j \simeq \bigotimes_{\ell=1}^{N} \rho_{j\ell}$.

The signal PSD is obtained from the two-point time correlation function, evaluated through the ensemble expectation value 
$\mathrm{Tr}[\rho_{jj'} \hat{D}_j \hat{D}_{j'}]$ for different $j$ and $j'$.  
For $j=j'$,
\begin{align}
\MyTr{\rho_j \hat{D}_j \hat{D}_j} = \frac{1}{N} \Tr[\rho_{j\ell} (\hat{d}_{j\ell})^2]+ \frac{N-1}{N} \left(\Tr[\rho_{j\ell} \hat{d}_{j\ell}]\right)^2.
\end{align}
Defining $\Tr[\rho_{j\ell} \hat{d}_{j\ell}]= T_{1j}$ and $\Tr[\rho_{j\ell} (\hat{d}_{j\ell})^2]= T_{2j}$, this becomes
\begin{align}
 \MyTr{\rho_j \hat{D}_j \hat{D}_j} = \frac{1}{N} T_{2j}+ \frac{N-1}{N} (T_{1j})^2.
\end{align}
For $j\neq j'$, the measurements are uncorrelated, so
\begin{align}
 \MyTr{(\rho_j \otimes \rho_{j'}) (\hat{D}_j \otimes \hat{D}_{j'})} = T_{1j}T_{1j'}.
\end{align}
Combining these, the two-point correlator can be approximated as
\begin{align}
 \MyTr{\rho_{jj'} \hat{D}_j \hat{D}_{j'}} \simeq \frac{1}{N} T_{2j}\delta_{jj'} + T_{1j}T_{1j'},
\end{align}
where the signal is contained in the second term. We define $\rho_{jj'}=\rho_j \otimes \rho_{j'}(\rho_j) $ for $j\neq j'(j=j')$ as in Ref.~\cite{Chigusa:2024psk} for qubits. For a qutrit, the signal component arises from correlations between the accumulated magnetic-field-induced phases,
$\langle \theta_{p j} \theta_{p j'} \rangle$.

Following the treatment in Ref.~\cite{Dror:2022xpi,Chigusa:2024psk}, we model the axion DM field as a coherent  monochromatic cosine wave with a stochastic random phase.  
The finite coherence time $\tau_a$ is incorporated by introducing a step-function cutoff
$\Theta(\tau_a - |t_j - t_{j'}|)$ in the correlation function.  
Under this approximation, the qutrit signal PSD is four times larger than that of a qubit, reproducing discussion of the main text.

We note that this simplified stochastic treatment neglects the detailed DM velocity distribution and instead imposes a sharp temporal cutoff to model coherence loss.  Further, this model
 omits Earth-induced modulations (e.g. daily or annual), which can become relevant for long integration times.  
More rigorous analyses that incorporate the velocity distribution have been considered in Refs.~\cite{Bloch:2022kjm,Lee:2022vvb,Wei:2023rzs,Xu:2023vfn}.  
Such refinements are not expected to qualitatively alter the projected sensitivity results and we defer comprehensive stochastic analysis for future work.

\subsection{C.2: Spin projection noise}

Here we evaluate the quantum spin-projection noise (SPN) associated with the measurements.  
The SPN power corresponds to the variance of the observable under the null hypothesis, that is in the absence of signal with $B_{\mathrm{eff}} = 0$.  
The auto-covariance of the measured observable is
\begin{align}
   \langle \Delta \hat{D}_j \Delta \hat{D}_{j'}\rangle &= \MyTr{\rho_{jj'} \hat{D}_j \hat{D}_{j'}} - \MyTr{\rho_j \hat{D}_j}\MyTr{\rho_{j'} \hat{D}_{j'}} \nonumber \\
   &= \frac{1}{N}\big(T_{2j}-T_{1j}^2\big)\delta_{jj'}\big|_{B_{\rm eff}=0}.
    \label{eq:qutrits_var}
\end{align}
where the Kronecker delta reflects that successive measurements are uncorrelated.

The corresponding noise PSD $\mathcal{B}_k$ is obtained as the discrete Fourier transform of this auto-covariance.  
Since the noise is uncorrelated between measurements $\propto \delta_{jj'}$, the PSD is flat (i.e. white noise) 
\begin{align}
    \mathcal{B}_k = \frac{\tau^2}{t_{\rm obs}}\sum_{j,j'}e^{2\pi i k (j-j')/N_{\rm obs}} \langle \Delta \hat{D}_j \Delta \hat{D}_{j'}\rangle = \frac{\tau}{N}\big(T_{2j}-T_{1j}^2)\big|_{B_{\rm eff}=0}. 
\end{align}
This expression reproduces the standard SPN limit derived for qubits in Ref.~\cite{Chigusa:2024psk}.

For the qutrit Ramsey sequence case, the measurement observable is $\hat{d}_{j\ell} = \hat{S}_z^2$.  
This choice is dictated by fluorescence readout, which cannot distinguish between the $\ket{+1}$ and $\ket{-1}$ states.  
The operator $\hat{S}_z^2$ conveniently assigns eigenvalue $1$ to the sensing subspace ($\ket{\pm1}$) and eigenvalue $0$ to the readout state ($\ket{0}$).  
For a spin-1 system, $(\hat{S}_z^2)^2 = \hat{S}_z^2$, implying  
\begin{equation}
    T_{2j} = \mathrm{Tr}[\rho_{j\ell}(\hat{S}_z^2)^2]
           = \mathrm{Tr}[\rho_{j\ell}\hat{S}_z^2]
           = T_{1j}.
\end{equation}

The Ramsey sequence operates at the point of maximum sensitivity (the ``mid-fringe''), corresponding to a zero-signal expectation value
\begin{align}
    T_{1j}\big|_{B_{\rm eff}=0} = T_{2j}\big|_{B_{\rm eff}=0} \simeq \frac{1}{2}.
\end{align}
A small signal $\theta_{p j}$ shifts this value as
$T_{1j} \simeq \tfrac{1}{2} - \theta_{p j}$.
The zero-signal variance per spin is therefore
\begin{align}
    \left(T_{2j}-T_{1j}^2\right)\big|_{B_{\rm eff}=0} = \frac{1}{2} - \left(\frac{1}{2}\right)^2 = \frac{1}{4}.
\end{align}
Consequently  the SPN PSD is $\mathcal{B}_k = \tau/(4N)$. 

This result is identical to that of a standard qubit Ramsey experiment~\cite{Chigusa:2024psk}, where the observable is $\hat{S}_z$  with eigenvalues $\pm1/2$, giving $T_{2j} = \langle \hat{S}_z^2 \rangle = 1/4$ and $T_{1j} = \langle \hat{S}_z \rangle = 0$, leading to the same variance of $1/4$.  
Since the qutrit signal PSD is four times larger while the SPN remains unchanged, the optimal qutrit sequence achieves an overall signal-to-noise-ratio (SNR) enhancement by a factor of 4.

\subsection{C.3: Photon shot noise}

The spin state of the NV center is determined by measuring its fluorescence, a process fundamentally limited by photon shot noise arising from statistical fluctuations in photon detection.  
This noise can significantly impact the overall magnetometric sensitivity.

Let $\alpha_{0}$ and $\alpha_{\pm}$ denote the average numbers of detected photons from the $\ket{S_z = 0}$ and $\ket{S_z = \pm1}$ states, respectively.  
These parameters incorporate both the emission probability and photon-collection efficiency.  
Both $\alpha_{0}$ and $\alpha_{\pm}$ increase with laser excitation power, which is assumed to be fixed in this analysis.  
For the detection of axion-electron interactions, the density matrix of a single NV-center electron spin is $\rho_{j\ell}$, while the corresponding density matrix of the emitted photon is
\begin{align}
  \rho_{j\ell}^{\mathrm{ph}} =&~ \Braket{0 | \rho_{j\ell} | 0} \left(
    (1-\alpha_0) \ket{0}_{j\ell}^{\mathrm{ph}} \bra{0} + \alpha_0 \ket{1}_{j\ell}^{\mathrm{ph}} \bra{1}
  \right) + \Braket{- |\rho _{j\ell} | -} \left(
    (1-\alpha_-) \ket{0}_{j\ell}^{\mathrm{ph}} \bra{0} + \alpha_- \ket{1}_{j\ell}^{\mathrm{ph}} \bra{1}
  \right)\nl
&+\Braket{+ | \rho_{j\ell} | +} \left(
    (1-\alpha_+) \ket{0}_{j\ell}^{\mathrm{ph}} \bra{0} + \alpha_+ \ket{1}_{j\ell}^{\mathrm{ph}}\right)
\end{align}
where $\ket{0}_{j\ell}^{\mathrm{ph}}$ and $\ket{1}_{j\ell}^{\mathrm{ph}}$ represent the states without and with photon detection, respectively.

For an $N$ ensemble of NV centers, the expected photon count is 
\begin{equation}
    \langle I_j \rangle^{\mathrm{ph}}
    = \mathrm{Tr}[\rho_j^{\mathrm{ph}} \hat{I}_j],
    \qquad
    \rho_j^{\mathrm{ph}}
    = \bigotimes_{\ell=1}^{N} \rho_{j\ell}^{\mathrm{ph}},
\end{equation}
where $\hat{I}_j = \sum_{\ell} \ket{1}_{j\ell}^{\mathrm{ph}} \bra{1}$.

For simplicity, we assume that all NV centers are aligned along one of the four crystallographic orientations that can be achieved using the techniques of Ref.~\cite{fukui2014perfect}. Without alignment, only one quarter of the NV centers are sensitive to the magnetic field, while the remaining three quarters contribute baseline fluorescence, increasing the overall shot noise.  
This results in an $\mathcal{O}(1)$ reduction in sensitivity relative to the aligned configuration.  Neglecting inhomogeneities that might cause dephasing, we obtain $\langle I_j \rangle^{\mathrm{ph}} = N n_j$, where the single-center expectation value is  
\begin{align}
  n_j = \Braket{+ | \rho_{j\ell} | +} \alpha_+ + \Braket{0 | \rho_{j\ell} | 0} \alpha_0+\Braket{- | \rho_{j\ell} |-} \alpha_-,
\end{align}
with an arbitrary choice of $\ell$.
The auto-covariance of the shot noise is then obtained as
\begin{align}
  \Braket{\delta I_j \delta I_{j'}}^{\mathrm{ph}}&=\MyTr{\rho_{jj'} \hat{I}_j \hat{I}_{j'}}-\MyTr{\rho_j \hat{I}_j}\MyTr{\rho_{j'} \hat{I}_{j'}}\Big|_{B_{\rm eff}=0}\nl
  &=\bigg( \Braket{I_j^2}^{\mathrm{ph}} - \left(
    \Braket{I_j}^\mathrm{ph}
  \right)^2\bigg)\delta_{jj'} = N^2 n_j (1-n_j)\delta_{jj'},
  \label{eq:shot_noise}
\end{align}
consistent with the expected binomial distribution result.  

Excluding the $k=0$ bin, which
 has a more complicated expression~\cite{Chigusa:2024psk} and is not of experimental relevance, 
the photon shot-noise PSD is
\begin{align}
    \mathcal{B}_{k\neq 0}^{\rm ph}=\tau N n_{\rm aver}(1-n_{\rm aver}),~~~~~n_{\rm aver}= \frac{1}{4}(\alpha_++\alpha_-)+\frac{1}{2}\alpha_0.
\end{align}

The corresponding sensitivity and its comparison to the SPN limit can be derived by restoring $B_{\rm eff}$ in Eq.~\ref{eq:shot_noise}, giving
\begin{align}
  \delta B_a^{\rm ph} = \left. \sqrt{\Var(\hat{I})} \left(
    \frac{d\Braket{I}}{d B_a}
  \right)^{-1} \right|_{B_a = 0},\nl
 \frac{d\Braket{I}/d B_a^{\rm ph}}{d\Braket{D}/d B_a} \simeq
  N [\alpha_0 - (\alpha_-+\alpha_+)/2],
  \label{eq:projection-shot-coefficients}
\end{align}
where $(\alpha_-+\alpha_+)/2 = \alpha_1$ recovers the same shot-noise result reported in Ref.~\cite{Chigusa:2024psk}.

To quantify the effect of imperfect photon collection, we can compare the ideal SPN-limited case with the more realistic case that includes photon shot noise.  
For nonzero Fourier modes with $k \neq 0$, the total sensitivity is degraded by a factor $\sigma_R$, defined as
\begin{equation}
  \sigma_R = \sqrt{1 + \frac{1}{C^2 n_{\rm aver}}}
  \label{eq:sigma_R_definition}
\end{equation}
where $C = (\alpha_0 - \alpha_1)/(\alpha_0 + \alpha_1)$ is the fluorescence contrast or fringe visibility~\cite{Barry:2019sdg,Barry:2023hon,taylor2008high}, and $n_{\mathrm{aver}}$ is the mean number of detected photons per readout. 
By definition, $\sigma_R = 1$ corresponds to a perfect readout limited only by fundamental SPN.

In practice, photon shot noise is often the dominant source of noise, yielding $\sigma_R \gg 1$.  
For example, typical bulk diamond measurements achieve $\sigma_R \sim 19$~\cite{balasubramanian2009ultralong,Shields:2014baz}, while optimized nano-beam geometries can improve this to $\sigma_R \simeq 10.6$~\cite{Shields:2014baz}.  
Including photon shot noise can thus degrade  the SPN-limited sensitivity of the main text by an additional factor $\sigma_R$, yet the qutrit-based setup still offers a substantial improvement over variety of existing terrestrial experiments. 

\section{D. Axion Dark Matter Sensitivity}

In an ensemble measurement, the fundamental sensitivity limit is determined by SPN.  
For an $N$ ensemble of  NV centers and a total of $N_{\mathrm{tot}}$ pulse sequence cycles during an observation time $t_{\mathrm{obs}}$, the signal uncertainty scales as $\Delta S_{\mathrm{sp}} \propto 1/\sqrt{N N_{\mathrm{tot}}}$.  
Assuming the SPN dominates, the minimum detectable magnetic field is
\begin{align}
    \Delta B_{\rm min} & \simeq \frac{e^{\tau/T_2^{*}}}{\gamma_e \sqrt{N \tau t_{\rm min}}|\Delta m|}   \simeq 1.45 {\rm fT}\left(\frac{10^{12}}{N}\right)^{1/2}\left(\frac{10 {\rm \mu s}}{\tau}\right)^{1/2}
    \left(\frac{1 {\rm s}}{t_{\rm min}}\right)^{1/2} \left(\frac{\sigma_R}{1}\right) ,
\label{eq:proj_sensitivity}
\end{align}
where $t_{\mathrm{min}} = \min(t_{\mathrm{obs}}, \tau_{\mathrm{DM}})$ accounts for the finite axion DM coherence time $\tau_{\mathrm{DM}}$, and the numerical value corresponds to a qutrit sensor with $|\Delta m|=2$.  
The factor $\sigma_R \ge 1$ quantifies degradation due to readout shot noise beyond the SPN limit.  
The sensitivity in Eq.~\eqref{eq:proj_sensitivity} when optimized at $\tau \simeq T_2^{*}/2$ scales as $\Delta B_{\mathrm{min}} \propto 1/(|\Delta m|\sqrt{T_2^{*}})$, underscoring the advantage of long coherence times and larger spin projection differences $|\Delta m|$ achievable in qutrit systems.

We now evaluate the SNR of PSD.  
For the signal PSD $\mathcal{S}_k$, we adopt the formulation of Ref.~\cite{Chigusa:2024psk} for qubits, appropriately modified by the qutrit enhancement factor.  
The total noise PSD $\mathcal{B}_k$  contains statistically independent contributions 
\begin{align}
    \mathcal{B}_k = \mathcal{B}_k^{\rm SPN} + \mathcal{B}_k^{\rm SN} + \mathcal{B}_k^{\rm others}.
\end{align}
Here $\mathcal{B}_k^{\mathrm{SPN}}$ represents SPN, $\mathcal{B}_k^{\mathrm{SN}}$ accounts for readout shot noise  and $\mathcal{B}_k^{\mathrm{others}}$ includes additional noise sources such as temperature drifts and electric field fluctuations~\cite{Barry:2019sdg}.  
In a controlled environment  $\mathcal{B}_k^{\mathrm{others}}$ is expected to be highly suppressed.

For frequencies $f_a \lesssim 1/\tau$ and observation times $t_{\mathrm{obs}} \lesssim \tau_a$ the SNR scaling behavior simplifies to
\begin{align}
    {\rm SNR} = \frac{ \mathcal{S}_k}{\mathcal{B}_k} \propto \begin{cases} \displaystyle \frac{\rho_{\rm DM}g_{aee}^2t_{\rm obs} \tau N |\Delta m|^2 \average{v_{\rm DM}^2}}{m_e^2 \sigma_R^2}, & {\rm for~Ramsey~sequence} \\
    \\
    \displaystyle \frac{\rho_{\rm DM}g_{aee}^2t_{\rm obs} m_a^2 \tau^3 N |\Delta m|^2\average{v_{\rm DM}^2}}{m_e^2 \sigma_R^2}, & {\rm for~Hahn-echo~sequence}
    \end{cases}
    \label{eq:snr_scaling}
\end{align}
where $\langle v_{\mathrm{DM}}^2 \rangle$ is the mean-squared DM velocity.  
The noise is dominated by SPN and photon shot noise both incorporated through $\sigma_R^2$.  
These scaling relations highlight the differing time dependencies of Ramsey and Hahn-echo protocols and quantitatively explain the improvement in sensitivity shown in main text Fig.~\ref{fig:ramsey_sensitivities}.

To estimate projected sensitivity, we employ the Asimov dataset methodology~\cite{Cowan:2010js}.  
Assuming that $\mathcal{S}_k$ and $\mathcal{B}_k$ follow exponential statistics, the profile log-likelihood test statistic~\cite{Dror:2022xpi,Lee:2022vvb,Chigusa:2024psk} is
\begin{align}
  q = -2\sum_{k} \left[
    \left(
      1 - \frac{\BPSD{k}}{\SPSD{k} + \BPSD{k}}
    \right)- \ln \left(
      1 + \frac{\SPSD{k}}{\BPSD{k}}
    \right)
  \right],
\end{align}
where $k$ indexes the frequency bins.  
The test statistic $q$ follows a chi-squared distribution with one degree of freedom, and the 95\% confidence level projected upper limit sensitivity on the signal strength is determined by solving $q = 2.71$.

\end{document}